\documentclass[10pt,conference]{IEEEtran}
\usepackage{epsfig,setspace,amsmath,epsf,amssymb,bm,theorem,cite,graphicx, epstopdf,algorithm,float,color,mathtools, authblk}
\usepackage[table,xcdraw]{xcolor}
\usepackage{subcaption}
\usepackage{algpseudocode}
\usepackage{bbm}
\usepackage[short,c2]{optidef}
\usepackage{physics}

\newtheorem{theorem}{Theorem}
\newtheorem{corollary}{Corollary}

\newtheorem{remark}{Remark}
\newtheorem{lemma}{Lemma}

\newcommand{\e}{{\mathbb{E}}}

\IEEEoverridecommandlockouts
\allowdisplaybreaks

\begin{document}

\title{Optimum Monitoring and Job Assignment with Multiple Markov Machines}

\author{Sahan Liyanaarachchi \qquad Sennur Ulukus\\
	\normalsize Department of Electrical and Computer Engineering\\
	\normalsize University of Maryland, College Park, MD 20742 \\
	\normalsize \emph{sahanl@umd.edu} \qquad \emph{ulukus@umd.edu}}

\maketitle

\begin{abstract}
We study a class of systems termed Markov Machines (MM) which process job requests with exponential service times. Assuming a Poison job arrival process, these MMs oscillate between two states, free and busy. We consider the problem of sampling the states of these MMs so as to track their states, subject to a total sampling budget, with the goal of allocating external job requests effectively to them. For this purpose, we leverage the \emph{binary freshness metric} to quantify the quality of our ability to track the states of the MMs, and introduce two new metrics termed \emph{false acceptance ratio} (FAR) and \emph{false rejection ratio} (FRR) to evaluate the effectiveness of our job assignment strategy. We provide optimal sampling rate allocation schemes for jointly monitoring a system of $N$ heterogeneous MMs.
\end{abstract}

\section{Introduction}
In any control process, timely estimation of the state of the system is crucial to make well informed decisions. To this end, age of information (AoI) has been a prominently used quantifier for timeliness \cite{age1, age2, yates2020age}. However, depending on the application, simply minimizing the staleness may be inadequate and quite often the impact of stale updates would depend on the actual state of the system as well. To address the drawbacks of AoI for such systems, age of incorrect information (AoII) was introduced in \cite{AoII2019} as a novel performance metric. There has been a variety of variations to AoII metric where the simplest of them being the binary freshness metric.

The binary freshness metric has been utilized in remote estimation applications involving Markovian sources due to its simplicity and its direct relation to the probability of error of the estimates\cite{melih_BF_cache, melih_IF_CUS,melih_BF_Inf, melih_BF_gossip}. However, as with AoI, depending on the application, its utility may be limited. In this regard, \cite{nail_QS} introduced two new variants of this metric termed \emph{fresh when close} (FWC) and \emph{fresh when sampled} (FWS), incorporating semantic relations between states of the system. Under these two new metrics, \cite{nail_QS} considers the problem of optimal sampling rate allocation for query based sampling of a system of heterogeneous Markov sources. A related work \cite{graves2024} considers the problem of monitoring distributed binary Markov sources and devises an update policy to minimize the probability of error of top-$k$ sources over a finite horizon.

In this work, we look at the optimal sampling rate allocation problem for query based sampling of a system of $N$ Markov machines (MMs), where each MM can be viewed as a binary Markov source. However, in this work, in addition to sampling, we consider that the monitor, i.e., the resource allocator (RA) in our case, makes decisions based on its current estimate of the system. These decisions can alter the state of the system and hence adds a new layer of complexity to the work in \cite{nail_QS}, where MMs had only internal dynamics. We consider a system where external job requests arrive at the RA following a Poison process and the RA either accepts or rejects these jobs based on its current estimate. If an accepted job gets assigned to the MM, the state of the MM will change. Hence, the decisions of the RA alters the state of the MM. For such systems, binary freshness alone, may not be an appropriate metric for the evaluation of the decisions made by the RA. Therefore, for such systems, we introduce two new metrics termed false acceptance ratio (FAR) and false rejection ratio (FRR) to quantify how the sampling rate affects the quality of decisions made by the RA.

In this paper, we first provide analytical expressions for the freshness, FAR and FRR metrics. Next, we compare how the three metrics are interrelated with each other and provide theoretical insight about them. Then, we consider a system of $N$ heterogeneous MMs and provide rate allocation policies to optimize these metrics. 

\begin{figure}
    \centering
    \includegraphics[scale=0.7]{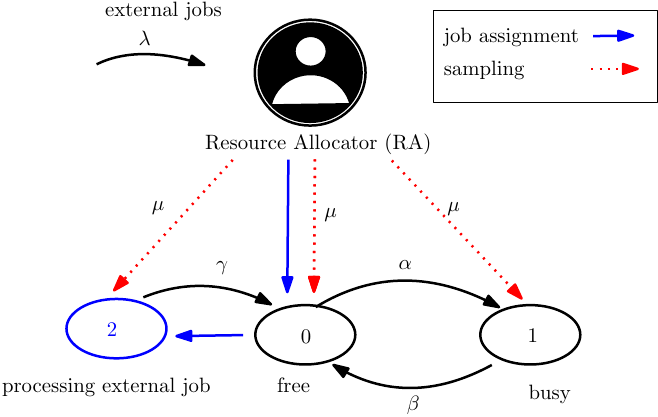}
    \caption{System model of a single MM: The MM oscillates between ``free'' (0) and ``busy'' (1) with rates $\alpha$ and $\beta$ due to internal jobs. When an external job is assigned, the state goes to ``process external job'' (2) and recovers to ``free'' with rate $\gamma$. MM is sampled with rate $\mu$. External jobs arrive with rate $\lambda$.}
    \label{fig:sys_model}
\end{figure}

\section{System model}
Let us represent our MM with a two state continuous time Markov chain (CTMC), where state $0$ represents the machine is free and state $1$ represents the machine is busy; see Fig.~\ref{fig:sys_model}. When the machine is free, let $\alpha$ denote the rate at which internal jobs arrive at the machine. These internal jobs will be completed by the machine at a rate of $\beta$. Let $X(t)$ denote the state of the MM at time $t$. A resource allocator will sample the MM at a rate of $\mu$, where we assume that once sampled, the RA will receive the information instantaneously. The RA will maintain an estimate of the state of MM denoted by $\hat{X}(t)$. 

External job requests arrive at the RA at a rate $\lambda$ and these job requests are allocated to the MM based on $\hat{X}(t)$. If an external job request is assigned to the MM, it will complete the job request with rate $\gamma$. Let $X(t)=2$ denote the state where the MM is processing (i.e., busy with) an external request. When an external job request arrives at the RA, it will reject the job request if $\hat{X}(t) \in \{1,2\}$ and will accept the job request if $\hat{X}(t)=0$. Once an external job request is accepted, RA will try to assign it to the MM. If $X(t)=0$, at this instance, the MM will immediately begin to process the external job request. However, if $X(t)\neq 0$ at this instance, the job will be lost to the RA. In either case, the act of assigning a job to the MM makes the RA be aware of the state of the MM and hence, RA can update its estimator at this time instance. Further, we assume that the initial state of the estimator and the MM are zero (free) state, i.e., $X(0)=\hat{X}(0)=0$. Next, we present the metrics of interest for our system model.

\subsection{Binary Freshness}
Here, we adopt the \emph{fresh when close} (FWC) definition for freshness introduced in \cite{nail_QS}. We say that our estimate is fresh when it is equal to the actual state in the semantic sense. Let $C(X(t),\hat{X}(t))\in \{0,1\}$ be a similarity, i.e., freshness, map between $X(t)$ and $\hat{X}(t)$. Then, the average freshness $\e[\Delta]$ is,
\begin{align}
    \e[\Delta]=\e\left[\limsup_{T\to \infty} \frac{1}{T}\int_{0}^{T} C(X(t),\hat{X}(t)) \,dt\right].
\end{align}
Depending on the setting, we may have different assignments for $C(X(t),\hat{X}(t))$. For instance, we may insist that our observation is fresh only when $\hat{X}(t)=X(t)$, in which case $C(0,0)=C(1,1)=C(2,2)=1$. Alternatively, we may consider the states 1 and 2 to be semantically identical, as they both represent ``busy'', in which case  $C(0,0)=C(1,1)=C(2,2)=C(1,2)=1$. This represents ``close'' in FWC.

\subsection{False Acceptance Ratio (FAR)}
Once an external job request arrives at the RA, the RA may choose to accept it or reject it based on its current estimate $\hat{X}(t)$ of the MM. However, if it chooses to accept a job when the MM is busy processing another job, this new job needs to be  discarded and hence the RA must be penalized. We term this event as a \emph{false acceptance}. Let $N_A(t)$ denote the total number of jobs the RA accepted by time $t$ and let $N_{FA}(t)$ denote the total number of false acceptances by time $t$. Then, 
\begin{align}
    \text{FAR}= \e\left[\limsup_{T\to\infty}\frac{N_{FA}(T)}{N_A(T)}\right].
\end{align}

\subsection{False Rejection Ratio (FRR)}
If the RA chooses to discard an external job request when the MM was actually free, we term this event as a \emph{false rejection}. Similar to the FAR, let $N_R(t)$ denote the total number of jobs the RA rejected by time $t$ and let $N_{FR}(t)$ denote the total number of false rejections by time $t$. Then, 
\begin{align}
    \text{FRR}= \e\left[\limsup_{T\to\infty}\frac{N_{FR}(T)}{N_R(T)}\right].
\end{align}
Note that the limit and expectation can be exchanged in all three metrics by the bounded convergence theorem \cite{koralov_sinai}. In general, higher $\e[\Delta]$ and lower FAR and FRR metrics are analogous to a good sampling strategy.

\begin{figure}
    \centering
    \includegraphics[scale=0.55]{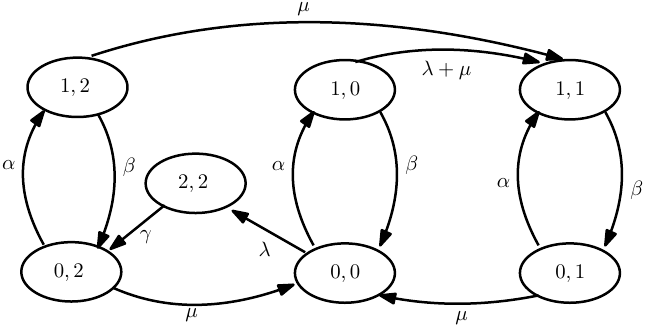}
    \caption{State transition diagram of $Y(t)$.}
    \label{fig:zero_buf_dsr}
\end{figure}

\begin{figure}
    \centering
    \includegraphics[scale=0.53]{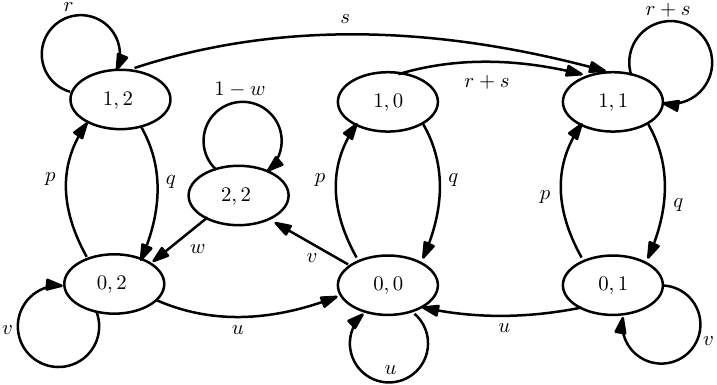}
    \caption{Jump chain of $Y(t)$.}
    \label{fig:zero_buf_ssr_jump}
\end{figure}

\section{Single Markov Machine}
To model the system with a single MM and a RA, we use a 2-dimensional Markov chain whose states are represented by  $Y(t)=(X(t),\hat{X}(t))\in S$ where $S = \{0,1\}\times\{0,1\}\cup \{0,1,2\}\times\{2\}$. Fig.~\ref{fig:zero_buf_dsr} represents the state transition diagram of this system. Since the Markov chain has a finite state space, a unique stationary distribution exists. Let $\bm\pi=\{\pi_i\}_{i\in S}$ denote the stationary distribution of $Y(t)$. Let $\bm Q$ denote the generator matrix of $Y(t)$ where $Q_{ij}$ is the rate of transition from state $i$ to state $j$ and $Q_{ii}=-\sum_j Q_{ij}$ with $i,j\in S$. Then, $\bm\pi$ satisfies $\bm\pi \bm Q=\bm0$. The binary freshness of the system can then be computed using $\bm \pi$ as,
\begin{align}
    \e[\Delta]=\sum_{(i,j)\in S} \pi_{ij} C(i,j).
\end{align}

Next, we evaluate the FAR and FRR metrics for this system. For this, we need to move from the CTMC to an equivalent discrete time Markov chain (DTMC) which is called the jump chain to model state space of the system based on the occurrences of events. Note that this is not the embedded DTMC of the CTMC, since the jump chain will involve the \emph{fictitious} self transitions. The jump chain of $Y(t)$ is given in Fig.~\ref{fig:zero_buf_ssr_jump} where $p=\frac{\alpha}{\mu+\lambda+\alpha}$, $u=\frac{\mu}{\mu+\lambda+\alpha}$, $v= \frac{\lambda}{\mu+\lambda+\alpha}$,  $q= \frac{\beta}{\mu+\lambda+\beta}$,  $r= \frac{\lambda}{\mu+\lambda+\beta}$, $s= \frac{\mu}{\mu+\lambda+\beta}$ and $w=\frac{\gamma}{\mu+\lambda+\gamma}$. Let $\tilde{\bm\pi}=\{\tilde{\pi}_i\}_{i\in S}$ be the stationary distribution of the jump chain of $Y(t)$. Then, Theorem~\ref{thrm:far_ffr} gives the expressions for FAR and FRR metric using the stationary distribution of the jump chain.

\begin{theorem} \label{thrm:far_ffr}
    Let $\mu>0$ and $\tilde{\bm \pi}=\{\tilde{\pi}_i\}_{i\in S}$ be the stationary distribution of the jump chain of the system and $p_i$ be the probability of transition in the jump chain when in state $i$ due to an external job arrival. Let $S_A$ and $S_{FA}$ denote  the sets of states in which the RA accepts jobs and  falsely accepts jobs, respectively. Let $S_R$ and $S_{FR}$ be the sets of states in which the RA rejects jobs and falsely reject jobs, respectively. Then, FAR and FRR metrics are given by,
    \begin{align}
        \text{FAR}=\frac{\sum_{i\in S_{FA}}p_i\tilde{\pi}_i}{\sum_{i\in S_A}p_i\tilde{\pi}_i}, \quad
        \text{FRR}=\frac{\sum_{i\in S_{FR}}p_i\tilde{\pi}_i}{\sum_{i\in S_R}p_i\tilde{\pi}_i}.
    \end{align}
\end{theorem}

Next, we present an important relation between the stationary distributions of the CTMC and the jump chain in Lemma~\ref{lem:jump_stat}.

\begin{lemma}\label{lem:jump_stat}
   Let  $\bm\pi$ be the stationary distribution of the CTMC and $\tilde{\bm\pi}$ be the stationary distribution of the jump chain. Let $\eta_i$ be the total rate of transition out of state $i$ including self transition rates. Then, $\tilde{\pi}_i \propto \eta_i\pi_i$.
\end{lemma}

Now, using Lemma~\ref{lem:jump_stat}, we can further simplify the expressions in Theorem~\ref{thrm:far_ffr} as shown in Corollary~\ref{cor:far_frr}.

\begin{corollary}\label{cor:far_frr}
     Let $\mu>0$ and $\bm\pi$ be the stationary distribution of the CTMC of the system. Then, FAR and FRR are given by,
    \begin{align}
        \text{FAR}=\frac{\sum_{i\in S_{FA}}\pi_i}{\sum_{i\in S_A}\pi_i},\quad
        \text{FRR}=\frac{\sum_{i\in S_{FR}}\pi_i}{\sum_{i\in S_R}\pi_i}.
    \end{align}
\end{corollary}

Note that job acceptance occurs in either states $(1,0)$ or $(0,0)$ where the jobs accepted in state $(1,0)$ constitute false acceptances. Thus, we have $S_A=\{(0,0),(1,0)\}$ while $S_{FA}=\{(1,0)\}$. Similarly, $S_R=\{(0,1),(1,1),(0,2),(1,2),(2,2)\}$ and $S_{FR}=\{(0,1),(0,2)\}$. Now, using Corollary~\ref{cor:far_frr} and $\bm\pi$, we write FAR and FRR metrics as,
\begin{align}
    \text{FAR}&=\frac{\pi_{10}}{\pi_{10}+\pi_{00}}=\frac{\alpha}{\mu+\kappa}, \label{eqn:far_gen}\\
    \text{FRR}&=\frac{\pi_{02}+\pi_{01}}{\pi_{02}+\pi_{01}+\pi_{12}+\pi_{11}+\pi_{22}}\\
    &=\frac{\gamma\beta(\mu(\lambda+\alpha)+\lambda \kappa)}{(\alpha\gamma+\lambda\beta)\mu^2+\tilde{\beta}\mu+\gamma\lambda(\beta+\alpha)\kappa}, \label{eqn:frr_gen}
\end{align}
where $\tilde{\beta}=(\lambda\beta(\lambda+\beta)+\gamma(\lambda\beta+\lambda\alpha+\alpha \kappa))$ and $\kappa=\lambda+\alpha+\beta$.

\begin{theorem}\label{thrm:mu_zero}
    When $\mu=0$, even though the recurrence of the chain breaks, since $X(0)=\hat{X}(0)=0$, the FAR and FRR can be obtained by taking limit as $\mu\to 0^{+}$ in \eqref{eqn:far_gen} and \eqref{eqn:frr_gen}. 
\end{theorem}

Next, we present some important insights about the three metrics under two special cases.

\subsection{Case 1: $C(1,2)=1$}\label{sec:sim_map}
Here, we  define the similarity map as $C(1,1)=C(1,2)=C(2,2)=C(0,0)=1$ and zero otherwise. Let us call this the special similarity map. Then, $\e[\Delta]$ reduces to the following,
\begin{align}
    \e[\Delta]=1-(\pi_{10}+\pi_{01}+\pi_{02}). \label{eqn:fresh_gen}
\end{align}
Let $\e[\tilde{\Delta}]=1-\e[\Delta]$ denote the expected staleness of our estimates. Then, the following relationship holds true,
\begin{align}
    \e[\tilde{\Delta}]&= \pi_{10}+\pi_{01}+\pi_{02}\leq \text{FAR}+\text{FRR}.\label{eqn:stale_bnd}
\end{align}
Thus, in here, FAR and FRR metrics provide an upper bound on the staleness of the estimator.

\begin{theorem}\label{thrm:mu_zero_age}
    Under the special similarity map, when $\mu=0$, $\e[\Delta]$ can be obtained by taking limit as $\mu\to 0^{+}$ in \eqref{eqn:fresh_gen}.
\end{theorem}

\subsection{Case 2: $\beta=\gamma$}
When $\beta=\gamma$, there is no value in the distinction between the states $(1,2)$ and $(2,2)$ with $(1,1)$, and $(0,2)$  with $(0,1)$. In this case, the FRR metric and $\e[\Delta]$ are given by,
\begin{align}
     \text{FRR}&=\frac{\beta}{\mu+\alpha+\beta}, \\
    \e[\Delta]&=\frac{\kappa\mu^2+(\kappa^2-2\alpha\beta)\mu+\lambda\alpha \kappa}{\kappa\mu^2+(\kappa^2+\beta\lambda)\mu+\lambda(\alpha+\beta)\kappa}.\label{eqn:fresh_spc}
\end{align}

\begin{remark}
    For $\mu>0$, when $\lambda \to 0^+$, $\e[\Delta]$ reduces to the expression obtained in \cite{nail_QS} for binary Markov sources.
\end{remark}

\begin{figure}
    \centering
    \includegraphics[scale=0.6]{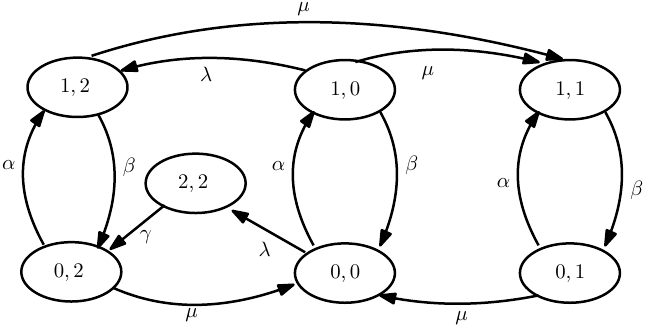}
    \caption{State transition diagram of a feedback free system.}
    \label{fig:no_ack}
\end{figure}

\begin{remark}
    When $\beta=\gamma$, the feedback free system (i.e., the MM does not report back whether it accepted a job or not) is equivalent to the system with feedback. The state transition diagram of the feedback free system is depicted in Fig.~\ref{fig:no_ack}
\end{remark}

Next, we highlight an important characteristic of the $\e[\tilde{\Delta}]$ in Lemma~\ref{lem:monotone}.

\begin{lemma}\label{lem:monotone}
    If $(\kappa-\alpha)^2+\alpha(\lambda-\alpha)\geq0$ then $\pdv{\e[\tilde{\Delta}]}{\mu} \leq 0$ for $\mu\geq 0$. Moreover, if $(\kappa-\alpha)^2+\alpha(\lambda-\alpha)<0$, then $\pdv{\e[\tilde{\Delta}]}{\mu}>0$ at $\mu=0$ and will eventually be negative as $\mu$ increases. 
\end{lemma}

\begin{remark}\label{rem:neg_samp}
Lemma~\ref{lem:monotone} indicates that freshness does not always increase with greater sampling rate and this counterintuitive phenomena can be viewed as a consequence of decision making with less informative data. This demonstrates that an ignorant decision maker, at times, can be better than an under informed decision maker.
\end{remark}

\begin{remark}
    When $\lambda>\alpha$, from Lemma~\ref{lem:monotone}, we have that $\e[\Delta]$ increases monotonically with $\mu$. Therefore, if the external job arrival rate is greater than the internal job arrival rate, then increasing the sampling rate improves freshness.
\end{remark}

\section{Multiple Markov Machines}
We consider a RA monitoring and allocating jobs for a system of $N$ MMs. Each MM is specialized for a single type of job and these jobs arrive at the RA at rates $\lambda_i$ for $1\leq i\leq N$. Assume that the $i$th MM is specialized for the $i$th job type. The RA samples the $i$th MM at a sampling rate of $\mu_i$ and we assume that the RA has a total sampling budget of $\Omega$. Let the internal job arrival rate of the $i$th MM be $\alpha_i$ and the job processing rate (external or internal) be $\beta_i$. Assume the similarity map in Section \ref{sec:sim_map} is enforced.

Let FAR$_i$, FRR$_i$ and $\e[\Delta_i]$ be the three metrics for the $i$th MM. Let us define the two weighted metrics, weighted action ratio (WAR) and the weighted average freshness (WAF) as,
\begin{align}
    \text{WAR}&= \sum_{i=1}^N w_i(w_A\text{FAR}_i+w_R\text{FRR}_i),\\
    \text{WAF}&= \sum_{i=1}^N w_i\e[\Delta_i],
\end{align}
where $w_A$, $w_R$, $w_i$ are nonnegative and satisfy $w_A+w_R=1$ and $\sum_{i=1}^N w_i=1$. Now, the problem at hand is to allocate the sampling rates $\mu_i$ to all MMs so as to optimize the above metrics. Since the sum of FAR and FRR can be used as a proxy for the staleness, we will develop schemes that optimize the two metrics separately and evaluate how one affects the other. 

\subsection{WAR Optimal Sampling}
In here, coincidentally, WAR must be minimized for better system performance. This problem is equivalent to solving the following convex optimization problem,
\begin{mini}
    {\mu_i\geq0}{\sum_{i=1}^N \frac{w_iw_A}{\mu_i+\kappa_i}+\frac{w_iw_R}{\mu_i+\alpha_i+\beta_i}}
    {\label{war_opt}}
    {}
    \addConstraint{\sum_{i=1}^N \mu_i}{\leq \Omega}.
\end{mini}
Define the Lagrangian of the above problem as,
\begin{align}
    L(\bm\mu,\bm\rho,\psi)= &\sum_{i=1}^N \frac{w_iw_A\alpha_i}{\mu_i+\kappa_i}+\frac{w_iw_R\beta_i}{\mu_i+\alpha_i+\beta_i}\nonumber \\
    &-\sum_{i=1}^N \rho_i\mu_i +\psi\Big(\sum_{i=1}^N \mu_i-\Omega\Big),
\end{align}
where $\bm\mu=\{\mu_i\}_{i=1}^N$ and $\rho_i$s and $\psi$ are nonnegative Lagrange multipliers. Since the optimization problem is convex and strictly feasible, it satisfies Slater's conditions. Then, the Karush-Kuhn-Tucker (KKT) conditions yield the following sufficient conditions for optimality \cite{boyd},
\begin{align}
    \psi&=\rho_i+\frac{w_iw_A\alpha_i}{(\mu_i+\kappa_i)^2}+\frac{w_iw_R\beta_i}{(\mu_i+\alpha_i+\beta_i)^2}, \label{eqn:grad}\\
    0&=\mu_i\rho_i, \label{eqn:slack}\\
    0&=\psi\Big(\sum_{i=1}^N\mu_i-\Omega\Big). \label{eqn:slack_2}
\end{align}
From \eqref{eqn:grad}, we have that $\psi>0$ and hence $\mu_i$s must satisfy $\sum_{i=1}^N\mu_i=\Omega$ by the virtue of \eqref{eqn:slack_2}. From \eqref{eqn:grad} and \eqref{eqn:slack}, we have that if $\mu_i>0$, then $\psi=\frac{w_iw_A\alpha_i}{(\mu_i+\kappa_i)^2}+\frac{w_iw_R\beta_i}{(\mu_i+\alpha_i+\beta_i)^2}$. Since both $\frac{1}{(\mu_i+\kappa_i)^2}$ and $\frac{1}{(\mu_i+\alpha_i+\beta_i)^2}$ are monotonically decreasing in $\mu_i$, for a fixed $\psi$ we can solve for $\mu_i$s using a bisection search. Since $\sum_{i=1}^N\mu_i$ decreases monotonically with $\psi$, we can employ another bisection search to find the optimal $\psi$. Hence, finding the optimal $\mu_i$ reduces to a water-filling algorithm. We refer to \cite{waterfilling}
for implementation details.
 
\subsection{WAF Optimal Sampling}
In here, the objective is to maximize freshness and this problem is equivalent to minimizing the staleness. Therefore, this problem reduces to the following optimization problem,
\begin{mini}
    {\mu_i\geq0}{\sum_{i=1}^N w_i\e[\tilde{\Delta}_i])}
    {\label{waf_opt}}
    {}
    \addConstraint{\sum_{i=1}^N \mu_i}{\leq \Omega},
\end{mini}
where $\e[\tilde{\Delta}_i]=1-\e[\Delta_i]$. Unlike the WAR metric, WAF is not always convex with respect to $\mu_i$s. However, since the problem satisfies the linear constraint qualification (LCQ), the KKT conditions provide the necessary conditions for optimality \cite{LIQC}. Let us define the Lagrangian of the problem in \eqref{waf_opt} as,
\begin{align}
    \tilde{L}(\bm\mu,\tilde{\bm\rho},\tilde{\psi})=&\sum_{i=1}^N\frac{\beta_i(\lambda_i+2\alpha_i)\mu_i+\lambda_i\beta_i\kappa_i}{\kappa_i\mu_i^2+(\kappa_i^2+\beta_i\lambda_i)\mu_i+\lambda_i(\alpha_i+\beta_i)\kappa_i}\nonumber\\
    &-\sum_{i=1}^N \tilde{\rho}_i\mu_i +\tilde{\psi}\Big(\sum_{i=1}^N \mu_i-\Omega\Big),
\end{align}
where $\tilde{\psi}$ and $\tilde{\rho}_i$s are the Lagrange multipliers. Then, the KKT conditions yield the following necessary conditions,
\begin{align}
    \tilde{\psi} =&\tilde{\rho}_i+\frac{\beta_i\kappa_i\left[(\lambda_i+2\alpha_i)\mu_i^2+2\lambda_i\kappa_i\mu_i\right]}{\left(\kappa_i\mu_i^2+(\kappa_i^2+\beta_i\lambda_i)\mu_i+\lambda_i(\alpha_i+\beta_i)\kappa_i\right)^2}\nonumber\\
    &+\frac{\beta_i\kappa_i\left[(\kappa_i-\alpha_i)^2+\alpha_i(\lambda_i-\alpha_i)\right]}{\left(\kappa_i\mu_i^2+(\kappa_i^2+\beta_i\lambda_i)\mu_i+\lambda_i(\alpha_i+\beta_i)\kappa_i\right)^2}, \label{eqn:fresh_grad}\\
    0=&\tilde{\rho_i}\mu_i, \label{eqn:fresh_slack_1}\\
    0=& \tilde{\psi}\Big(\sum_{i=1}^N\mu_i-\Omega\Big). \label{eqn:fresh_slack_2}
\end{align}

Next, we observe the following property of this problem.

\begin{lemma}\label{lem:waf_opt}
   If either $\max_i(\kappa_i-\alpha_i)^2+\alpha_i(\lambda_i-\alpha_i)>0$ or $\max_i(\lambda_i-\alpha_i)>0$, then the optimal $\mu_i$s must satisfy $\sum_{i=1}^N\mu_i=\Omega$.
\end{lemma}

Let us now assume that for at least one MM, we have $\lambda_i>\alpha_i$. Therefore, from Lemma~\ref{lem:waf_opt}, our problem reduces to the optimization problem in \eqref{waf_opt} where the inequality in the constraint is replaced with an equality as,
\begin{align}
    \sum_{i=1}^N \mu_i= \Omega. \label{waf_opt_2}
\end{align}

Now, note that the modified optimization problem is solved on a scaled simplex given in \eqref{waf_opt_2}. Denote this scaled simplex by $\mathcal{S}_{\Omega}$. To solve this optimization problem, we propose to use the projected gradient descent\cite{recht_wright}. Let $h(\bm \mu)$ be objective of \eqref{waf_opt}. The projected gradient descent reduces to the following iterative update given by  $\bm \mu^{(k+1)}=\textbf{Proj}_{\mathcal{S}_{\Omega}}(\bm \mu^{(k}-\tau \nabla h(\bm\mu^{(k)}))$ where $\tau$ is the learning rate and $\textbf{Proj}_{\mathcal{S}_{\Omega}}$ is the projection of a vector on to $\mathcal{S}_{\Omega}$ in \eqref{waf_opt_2}. Projection on to the standard simplex  has been studied and we adopt the algorithm presented in \cite{simplex_proj} for our problem. For $L$-smooth functions (see Lemma~\ref{lem:Lip}) the projected gradient descent is known to converge to a stationary point for a sufficiently small step size $\tau<\frac{1}{L}$ \cite{recht_wright}. However, since the problem is non-convex, the solution may not be globally optimal (possibly a local minima or a saddle point). Therefore, we run the algorithm several times with different initializations (from each vertex and the center of the scaled simplex along with some random initializations on $S_{\Omega}$) and prune out the best possible solution.

\begin{lemma}\label{lem:Lip}
   There exists a $L>0$ such that $\nabla h(\bm\mu)$ is $L$-Lipschitz on the nonnegative orthant. Alternatively, $h(\bm\mu)$ is $L$-smooth on the nonnegative orthant.
\end{lemma}

\section{Numerical Results}
\begin{figure}[t]
    \centering
    \begin{subfigure}[b]{\columnwidth}
         \centering
         \includegraphics[scale=0.5]{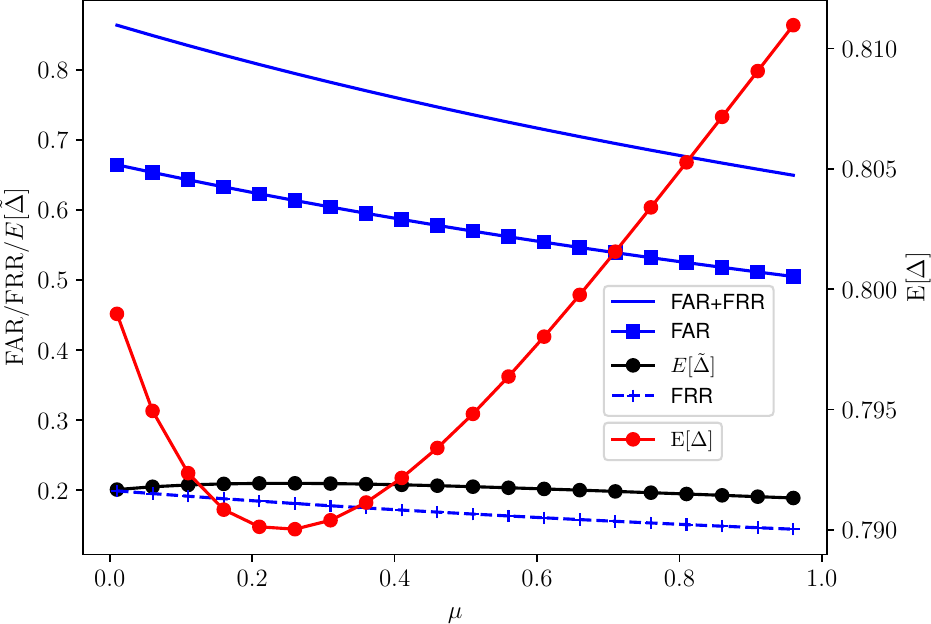}
         \caption{ $\alpha=2$.}
         \label{fig:var_mu_1}
    \end{subfigure}
    \begin{subfigure}[b]{\columnwidth}
        \vspace{5mm}
         \centering
         \includegraphics[scale=0.5]{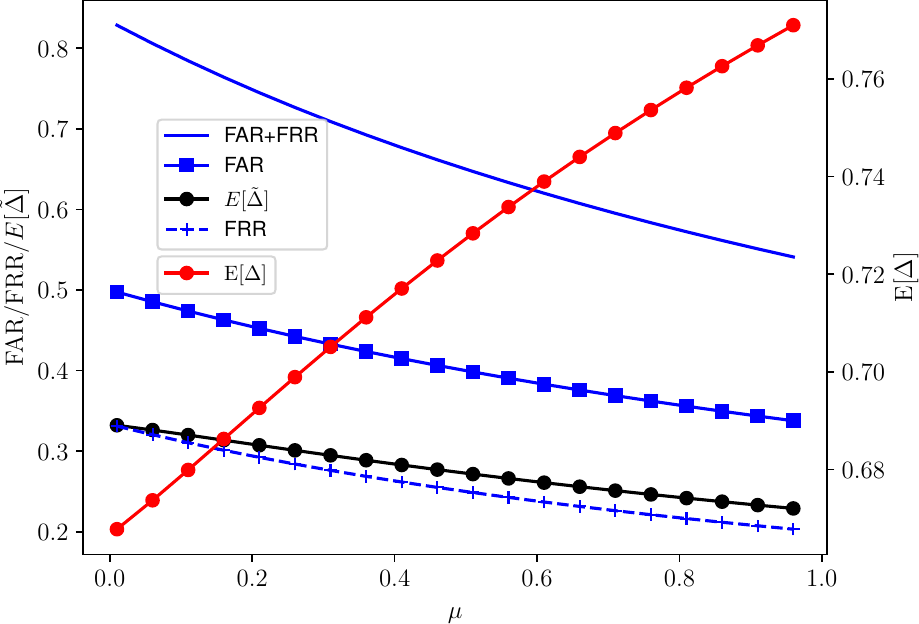}
         \caption{$\alpha=1$}
         \label{fig:var_mu_2}
    \end{subfigure}
    \caption{Variation of $\e[\Delta]$, $\text{FAR}$ and $\text{FRR}$ with $\mu$ for $\lambda=\beta=\gamma=0.5$. }
    \label{fig:var_mu}
\end{figure}

In this section, we evaluate and compare how the $\e[\Delta]$, FAR and FRR are related to each other. In all experiments, we assume that $\beta=\gamma$ and the similarity map is as described in Section \ref{sec:sim_map}. In the first experiment, we illustrate the variation of the above metrics with $\mu$ for two particular scenarios where in the first scenario, we set the rates such that they satisfy $(\kappa-\alpha)^2+\alpha(\lambda-\alpha)<0$ and in the second scenario this inequality is violated. As depicted in Fig.~\ref{fig:var_mu}, in the first scenario (see Fig.~\ref{fig:var_mu_1}) as $\mu$ increases, $\e[\Delta]$  first decreases and then increases. This illustrates the seemingly counterintuitive phenomena highlighted in Remark~\ref{rem:neg_samp}. On the other hand, in the second scenario (see Fig.~\ref{fig:var_mu_2}), $\e[\Delta]$ increases monotonically with $\mu$ as a consequence of Lemma \ref{lem:monotone}. In either case, as indicated by \eqref{eqn:stale_bnd}, staleness ($\e[\tilde{\Delta}]$) is upper bounded by the sum of FAR and FRR.

\begin{figure}[t]
    \centering
    \includegraphics[scale=0.5]{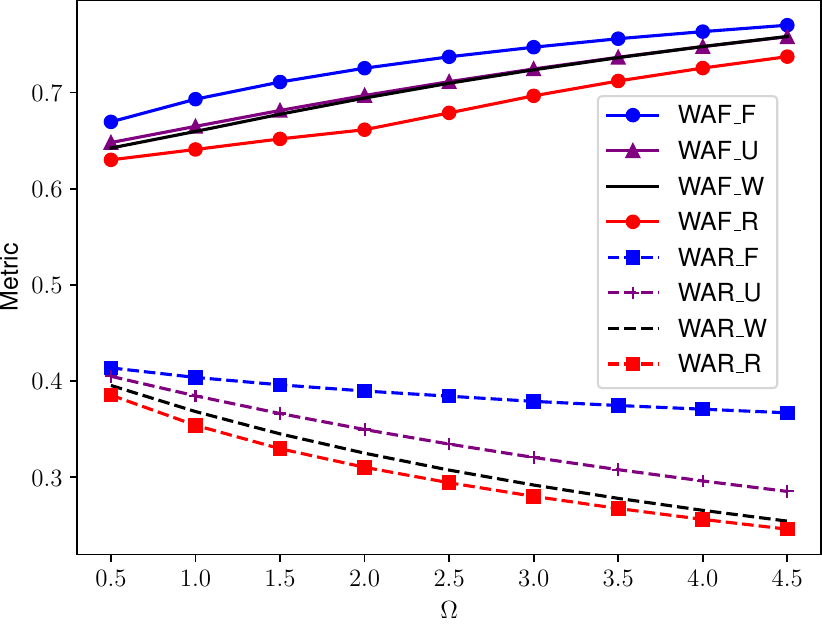}
    \caption{Variation of WAR and WAF with $\Omega$  under different sampling strategies for $\bm \alpha=\{2,1,1\}$, $\bm \beta = \{0.5,2,1.5\}$, $\bm \lambda=\{0.5,1.5,4\}$, $\bm w =\{0.6,0.1,0.3\}$, $w_A=0.6$, and $w_R=0.4$.}
    \label{fig:war_v_waf}
\end{figure}

In the next experiment, we consider a system of $N=3$ MMs and illustrate how the WAR and WAF metrics behave under different sampling policies/strategies. Let us denote the two metrics under a WAR optimal sampling policy as WAR$\_$R,  WAF$\_$R and under a WAF optimal sampling policy as  WAR$\_$F,   WAF$\_$F. Let WAR$\_$U,  WAF$\_$U and WAR$\_$W,  WAF$\_$W  be the same two metrics under a naive uniform sampling policy (i.e., $\mu_i=\frac{\Omega}{N}$) and a weight based sampling policy (i.e., $ \mu_i=w_i\Omega$). Fig.~\ref{fig:war_v_waf} illustrates the variation of the above metrics with $\Omega$. As seen in Fig.~\ref{fig:war_v_waf}, WAR$\_$R outperforms all other policies in terms of WAR metric and similarly WAF$\_$F outperforms all other policies in terms of freshness. Therefore, depending on the system preference,
the appropriate metric and the rate allocation policy must be carefully chosen.

\section{Conclusion}
In this work, we studied the problem of optimal sampling rate allocation for monitoring multiple Markov machines. In this regard, we introduced two new metrics to quantify the effect of the sampling rate on the RA's job assignment decisions. We evaluated our system performance from a freshness perspective (WAF) and as well as from a decision perspective (WAR), and showed how they differ from each other. We provided the optimal sampling rates for the minimization of the WAR metric and provided an algorithm to find locally optimal sampling rates to maximize WAF metric. Future research directions include the incorporation of a job queue or considering a single shared job process for all MMs.

\appendices
\section{Proof of Theorem \ref{thrm:far_ffr}}
Let $N_{A,i}(t)$ denote the total number of job acceptances when in state $i$ by time $t$. Let $N_i(T)$ denote total number of visits to state $i$ by time $T$   in the CTMC including the \emph{fictitious} self transitions as well (this corresponds to the self transitions in the jump chain). Let $N(T)$ be total number of transitions (including self transitions) in the CTMC by time $T$ . Then the following holds true,
    \begin{align}
        \limsup_{T\to\infty}\frac{N_{FA}(T)}{N_A(T)}&=\limsup_{T\to\infty}\frac{\sum_{i\in S_{FA}} N_{A,i}(T)}{\sum_{i\in S_A}N_{A,i}(T)}\\
        &=\limsup_{N(T)\to\infty}\frac{\sum_{i\in S_{FA}} \frac{N_{A,i}(T)}{N_i(T)}\frac{N_i(T)}{N(T)}}{\sum_{i\in S_A}\frac{N_{A,i}(T)}{N_i(T)}\frac{N_i(T)}{N(T)}}\\
        &=\frac{\sum_{i\in S_{FA}}p_i\tilde{\pi}_i}{\sum_{i\in S_A}p_i\tilde{\pi}_i}, \quad a.s\label{eqn:jmp_conv}
    \end{align}
    In here, \eqref{eqn:jmp_conv} holds almost surely ($a.s$) and was obtained using the fact, $N_i(T)\to\infty$ since the jump chain is recurrent ($\mu>0$) and hence $\frac{N_{A,i}(T)}{N_i(T)}\to p_i$ while $\frac{N_i(T)}{N(T)}\to\tilde{\pi}_i$. The expression for FRR is proved similarly.
\section{Proof of Lemma \ref{lem:jump_stat}}
    Let $\bm\Lambda$ be the transition matrix of the jump chain. Since $\bm\pi$ is the stationary distribution of the CTMC, we have,
    \begin{align}
        -Q_{i,i}\pi_i&= \sum_{j\neq i} Q_{j,i}\pi_j\\
        -\frac{Q_{i,i}}{\eta_i}\eta_i\pi_i&=\sum_{j\neq i}\frac{Q_{j,i}}{\eta_j}\eta_j\pi_j\\
        (1-\Lambda_{i,i})\eta_i\pi_i&=\sum_{j\neq i} \Lambda_{j,i}\eta_j\pi_j\\
        \eta_i\pi_i &= \sum_j\Lambda_{j,i}\eta_j\pi_j
    \end{align}
    Therefore, $\tilde{\pi_i}=\eta_i\pi_i$ satisfies $\bm\tilde{\pi}\bm\Lambda=\bm\tilde{\pi}$ and hence normalizing it yields the stationary distribution of the jump chain.
\section{Proof of Corollary \ref{cor:far_frr}}
    Let $\eta_i$ be the total rate transition out of state $i$ including self transitions in the CTMC. Let $\tilde{\pi}$ be the stationary distribution of the jump chain and $p_i$ be the probability of transition due to an external job arrival when in state $i$ of the jump chain. Then FAR metric is given by,
    \begin{align}
         \text{FAR}&=\frac{\sum_{i\in S_{FA}}p_i\tilde{\pi}_i}{\sum_{i\in S_A}p_i\tilde{\pi}_i}\\
         &=\frac{\sum_{i\in S_{FA}}p_i\eta_i\pi_i}{\sum_{i\in S_A}p_i\eta_i\pi_i}\\
         &=\frac{\sum_{i\in S_{FA}}\lambda\pi_i}{\sum_{i\in S_A}\lambda\pi_i}\label{eqn:eta_pi}\\
         &=\frac{\sum_{i\in S_{FA}}\pi_i}{\sum_{i\in S_A}\pi_i}
    \end{align}
In here, \eqref{eqn:eta_pi} was obtained using the fact that $p_i\eta_i=\lambda$. A similar proof follows for the FRR metric.
\section{Proof of Theorem \ref{thrm:mu_zero} $\&$ \ref{thrm:mu_zero_age}}
Note that when $\mu=0$, the states $(0,0)$ and $(1,1)$ will be transient and the chain will exit this transient state pair only upon an arrival of an external job request. Starting from $(0,0)$, the chain will exit from $(0,0)$ with probability $\frac{\lambda+\beta}{\lambda+\alpha+\beta}$ and from $(1,0)$ with probability $\frac{\alpha}{\lambda+\alpha+\beta}$. If the chain exits from $(0,0)$, the chain will eventually be in state $(1,2)$ with probability $\frac{\alpha}{\alpha+\beta}$ and on state $(0,2)$ with probability $\frac{\beta}{\alpha+\beta}$. Similarly, if the chain exited from $(1,0)$, it will eventually be in state $(1,1)$ with probability $\frac{\alpha}{\alpha+\beta}$ and in state $(0,1)$ with probability $\frac{\beta}{\alpha+\beta}$. If the chain exited the transient state pair from $(0,0)$, then  FAR  will be zero and if it exited from $(1,0)$ it will be $1$. In both cases, the FRR metric will be $\frac{\beta}{\alpha+\beta}$. Therefore, when $\mu=0$, the FAR and FRR metrics will be given by,
\begin{align}
    \text{FAR}&= 0.\frac{\lambda+\beta}{\lambda+\alpha+\beta}+1.\frac{\alpha}{\lambda+\alpha+\beta}=\frac{\alpha}{\kappa}\\
    \text{FRR}&=\frac{\beta}{\alpha+\beta}.\frac{\lambda+\beta}{\lambda+\alpha+\beta}+\frac{\beta}{\alpha+\beta}.\frac{\alpha}{\lambda+\alpha+\beta}
    &=\frac{\beta}{\alpha+\beta}
\end{align}
Similarly, under the special similarity map, it can be shown that $\e[\Delta]=\frac{\alpha}{\alpha+\beta}$ regardless the state from which the chain exited the transient state pair.
\section{Proof of Lemma \ref{lem:monotone}}
From \eqref{eqn:fresh_spc}, we have that,
\begin{align}
    \e[\tilde{\Delta}]= \frac{\beta(\lambda+2\alpha)\mu+\lambda\beta\kappa}{\kappa\mu^2+(\kappa^2+\beta\lambda)\mu+\lambda(\alpha+\beta)\kappa}
\end{align}
This gives, $\pdv{\e[\tilde{\Delta}]}{\mu}$  as,
\begin{align}
\pdv{\e[\tilde{\Delta}]}{\mu}&=-\frac{\beta\kappa\left[(\lambda+2\alpha)\mu^2+2\lambda\kappa\mu\right]}{\left(\kappa\mu^2+(\kappa^2+\beta\lambda)\mu+\lambda(\alpha+\beta)\kappa\right)^2}\nonumber\\
    &\qquad-\frac{\beta\kappa\left[(\kappa-\alpha)^2+\alpha(\lambda-\alpha)\right]}{\left(\kappa\mu^2+(\kappa^2+\beta\lambda)\mu+\lambda(\alpha+\beta)\kappa\right)^2}
\end{align}
Clearly, the first term is negative for all $\mu>0$ while the second term is negative only if $(\kappa-\alpha)^2+\alpha(\lambda-\alpha)\geq 0$. If $(\kappa-\alpha)^2+\alpha(\lambda-\alpha)<0$, then at $\mu=0$, $\pdv{\e[\tilde{\Delta}]}{\mu}>0$ and it will decrease below zero as $\mu$ increases. This proves the result.
\section{Proof of Lemma \ref{lem:waf_opt}}

From \eqref{eqn:fresh_grad}, we have that if at least for one MM, $(\kappa_i-\alpha_i)^2+\alpha_i(\lambda_i-\alpha_i)>0$, then $\tilde{\psi}>0$ and therefore from \eqref{eqn:fresh_slack_2}, we have $\sum_{i=1}^N\mu_i=\Omega$ for this case. When $\lambda_i>\alpha_i$, the above conditioned is satisfied trivially.
\section{Proof of Lemma \ref{lem:Lip}}
Since $h(\bm\mu)$ is linearly separable with respect to $\mu_i$s, it is sufficient to show that $\dv{\e[\tilde{\Delta_i}]}{\mu_i}$ is $L$-Lipschitz for $\mu_i\geq 0$. To show this, it is sufficient to show that $|\dv[2]{\e[\tilde{\Delta_i}]}{\mu_i}|\leq L$ for $\mu_i\geq 0$.
For brevity, let us denote $\e[\tilde{\Delta_i}]=\frac{l(\mu_i+a)}{\mu_i^2+b\mu_i+c}$ where $a$,$b$,$c$ and $l$ are all positive constants. Then $\dv[2]{\e[\tilde{\Delta_i}]}{\mu_i}$ will be given by,
\begin{align}
    \dv[2]{\e[\tilde{\Delta_i}]}{\mu_i}=\frac{2l(ab^2-ac-bc+3ab\mu_i-3c\mu_i+3a\mu_i^2+\mu_i^3)}{(\mu_i^2+b\mu_i+c)^2}
\end{align}
Clearly, as $\mu_i\to\infty$, we have $ \dv[2]{\e[\tilde{\Delta_i}]}{\mu_i}\to 0$. Therefore $\exists \mu_{L_i}>0$ such that $|\dv[2]{\e[\tilde{\Delta_i}]}{\mu_i}|<1, \quad\forall \mu_i>\mu_{L_i}$. Now, since $\dv[2]{\e[\tilde{\Delta_i}]}{\mu_i}$ is continuous on the compact interval $[0,\mu_{L_i}]$, it is bounded. Therefore $\exists L_i>0$ such that $|\dv[2]{\e[\tilde{\Delta_i}]}{\mu_i}|<L_i$. Now, let $L=\max_i L_i$. 

\bibliographystyle{unsrt}
\bibliography{refs}

\end{document}